\documentclass[%
superscriptaddress, %
twocolumn, %
showpacs,
 amsmath,amssymb,
 aps,
prb,
floatfix, longbibliography ]{revtex4-1}
\usepackage{graphicx}
\usepackage{dcolumn}
\usepackage{bm}
\usepackage{soul,color}

\usepackage[colorlinks,linkcolor=blue,anchorcolor=blue,citecolor=blue,urlcolor=blue]{hyperref}
\usepackage{tabularx}
\usepackage{mathtools}

\newcommand{\mb}[1]{\mathbf{#1}}

\begin{document}

\title{Emergence of intrinsically isolated flat bands and their topology in fully relaxed twisted multi-layer graphene}

\author{Xianqing Lin}
\email[E-mail: ]{xqlin@zjut.edu.cn}
\affiliation{College of Science,
             Zhejiang University of Technology,
             Hangzhou 310023, People's Republic of China}

\author{Haotian Zhu}
\affiliation{College of Science,
             Zhejiang University of Technology,
             Hangzhou 310023, People's Republic of China}

\author{Jun Ni}
\affiliation{State Key Laboratory of Low-Dimensional Quantum Physics and Frontier Science Center for Quantum Information,
             Department of Physics, Tsinghua University, Beijing 100084,
             People's Republic of China}

\date{\today}

\begin{abstract}
We study the electronic structure and band topology of fully relaxed
twisted multi-layer graphene (TMLG).
Isolated flat bands emerge in TMLG with the number of layers [$M+N$ with $M$ the layer number of the bottom few-layer graphene (FLG)]
up to 10
and with various stacking orders,
and most of them are on the hole side.
The touched bands of FLGs around the Fermi level
are split by the moir\'{e} coupling through the electron-hole asymmetry in low-energy bands of FLGs and by
the vertical hopping between next-nearest layers.
The full structural relaxation leads to global gaps that completely isolate a flat band.
For TMLG with given $M$ and $N$, the highest magnitude of Chern numbers ($|C|$) of the separable flat bands
reaches $M+N-1$ and can be hosted by certain
isolated bands.
The $|C|=9$ occurs in the isolated flat valence band of several configurations with 10 layers.
Such high $|C|$ originates from the lifting of the band-state degeneracy in the weak regime of
moir\'{e} coupling or from the topological phase transitions induced by the strong
moir\'{e} coupling.
Moreover, large orbital magnetic moments arise in isolated flat bands with high $|C|$ and
depend on the structural configurations of TMLG.
\end{abstract}

\pacs{%
}



\maketitle


\section{Introduction}

The emergence of low-energy flat bands
\cite{Bistritzer12233,LopesdosSantos2012,Fang2016,OriginPhysRevLett.122.106405} and
the observation of associated
superconductivity and correlated-insulator
phases\cite{Cao2018,cao2018unconventional,lu2019superconductors,xie2019spectroscopic,Maximized-electron-2019,Charge-order-2019,uri2020mapping,Strongly-correlated-2020}
in magic-angle twisted bilayer graphene (TBG) have inspired
great interest in exploring the peculiar
electronic structure of graphene moir\'{e} systems\cite{Graphene-bilayers-2020,The-marvels-2021}.
Finite Berry phases around the Dirac cones of monolayer graphene are maintained in
TBG\cite{All-Magic-Angles,Faithful-tight-binding,Failure-of-2019,Pseudo-Landau-2019}, while the breaking of $C_2T$ symmetry is required to obtain separable flat bands
with nontrivial Chern numbers ($C$)\cite{Mechanism2020nick,Twisted209zhang,Anomalous-Hall-2020}, which can be achieved by
carefully aligning TBG with hexagonal boron nitride to commensurate
twist angles between them\cite{Symmetry-breaking-2020,Band-structure-2020,Misalignment2021Lin,Moire2021Shi,Quasiperiodicity2021Mao}.
In such heterostructures, quantized anomalous Hall conductivity (QAHC) $\sigma_{xy} = C e^2/h$ is observed
at an odd filling of a flat band together with spontaneous orbital ferromagnetism\cite{Intrinsic2020Serlin,EmergentSharpe605}.
QAHC with $|C|=2$ has been realized in chirally stacked trilayer graphene aligned with BN\cite{Tunable-correlated-2020}.
In contrast to TBG, intrinsic flat Chern bands in a valley arise in BA-AB stacked twisted double bilayer graphene (TDBG)
and their $|C|$ can reach 3\cite{Theory-of-2019,Nearly-flat-2019,Quantum2019Liu,Band2019Koshino,Flat2019Chebrolu,Pressure2020lin}.
This suggests that higher-order flat Chern bands may occur in thicker twisted multi-layer graphene (TMLG)
composed of two few-layer graphene (FLGs) with relative rotation.
Moreover, complete isolation of a flat Chern band by global gaps from other bands is also essential to obtain QAHC that is contributed
purely by edge states\cite{Moire2021Shi}.
Therefore, it is important to explore TMLG systematically to identify isolated flat bands with high-order topology in graphene
moir\'{e} systems.

The electronic structure of TMLG depends
on the stacking orders of the FLGs and the twist angle ($\theta$)
\cite{Electronic2018Vela,Quantum2019Liu,Twists2019Cea,Electronic2020Georgios,zhang2020chiral,cao2020textit,ma2020moir}.
The flat bands around the Fermi level ($E_F$)
of TMLG were found to be entangled with each other or
with other dispersive bands by band crossings\cite{Quantum2019Liu,Electronic2020Georgios,zhang2020chiral,cao2020textit,ma2020moir}.
Most studies introduced external electric field to separate the flat bands so that their
band topology becomes well defined, while the
produced $C$ is limited to small values\cite{zhang2020chiral,ma2020moir}.
For TMLG composed of chirally stacked FLGs, the flat valence and conduction bands in a valley can be separated from other bands and their
total $C$ were demonstrated to increase with the layer numbers of TMLG\cite{Quantum2019Liu}.
This suggests that high $C$ may occur in one flat band if it can become isolated.
We note that only rigid moir\'{e} superlattices of TMLG were considered in these previous studies
\cite{Quantum2019Liu,Electronic2020Georgios,zhang2020chiral,cao2020textit,ma2020moir}, while
full relaxation has been shown to be able to enhance the band separation in TBG and TDBG
\cite{Nam2017,Atomicyoo2019atomic,CrucialPhysRevB.99.195419,ContinuumPhysRevB.99.205134,IntrinsicPhysRevB.100.201402,Pressure2020lin}.
In addition, the interlayer coupling was limited to that between adjacent layers
\cite{Quantum2019Liu,Electronic2020Georgios,zhang2020chiral,cao2020textit,ma2020moir}.
The coupling between next-nearest layers may also play an important role to isolate the flat bands.

Here, we have identified various stacking orders of fully relaxed TMLG with isolated flat bands.
Certain isolated flat valence bands in configurations with $M+N$ layers can host $|C|$ as large as $M+N-1$
and also large orbital magnetic moments for $M+N$ up to 10.
The mechanism behind the emergence of isolated flat bands and their significant band topology have been revealed.

\begin{figure*}[t]
\begin{center}
\includegraphics[width=1.9\columnwidth]{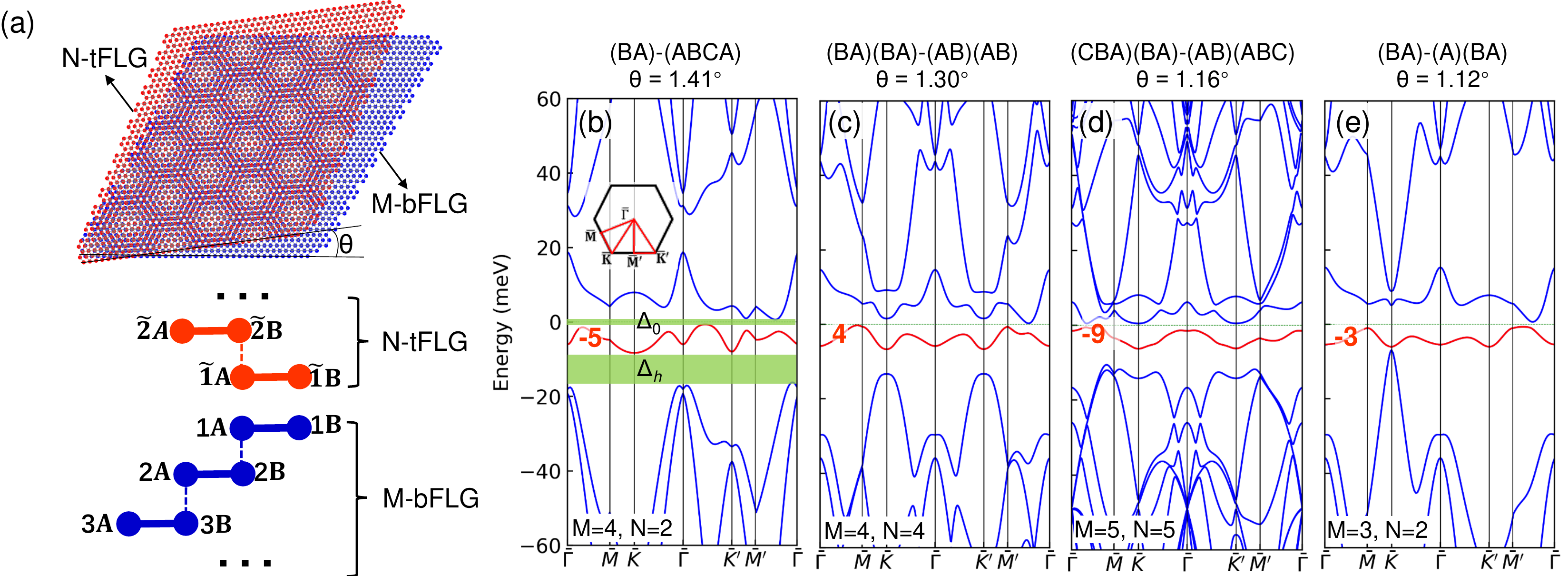}
\end{center}
\caption{(Color online) The geometry and electronic band structures of TMLG. (a)
The schematic view of the TMLG with the top FLG (tFLG) rotated
by $\theta$ counterclockwise with respect to the bottom FLG (bFLG).
The shown side view is around the AA stacking between layers $1$ and $\tilde{1}$.
The layer numbers of bFLG and tFLG are denoted by $M$ and $N$, respectively.
The sublattices in each layer are labeled by $n\alpha$ with $n$ the layer index and $\alpha$=A, B.
(b-e) The band structures with isolated flat valence bands (red lines) for typical configurations of the four stacking types of TMLG.
As bands in the two valleys are related by the time reversal symmetry,
only bands in the $\xi = +$ valley are displayed.
The isolated valence band is separated from other bands by global gaps $\Delta_0$ and $\Delta_h$ labeled in (b).
The Chern numbers of the isolated bands in the $\xi = +$ valley are labeled.
The Fermi levels are represented by dashed lines.
\label{fig1}}
\end{figure*}

\begin{figure}[t]
\begin{center}
\includegraphics[width=0.8\columnwidth]{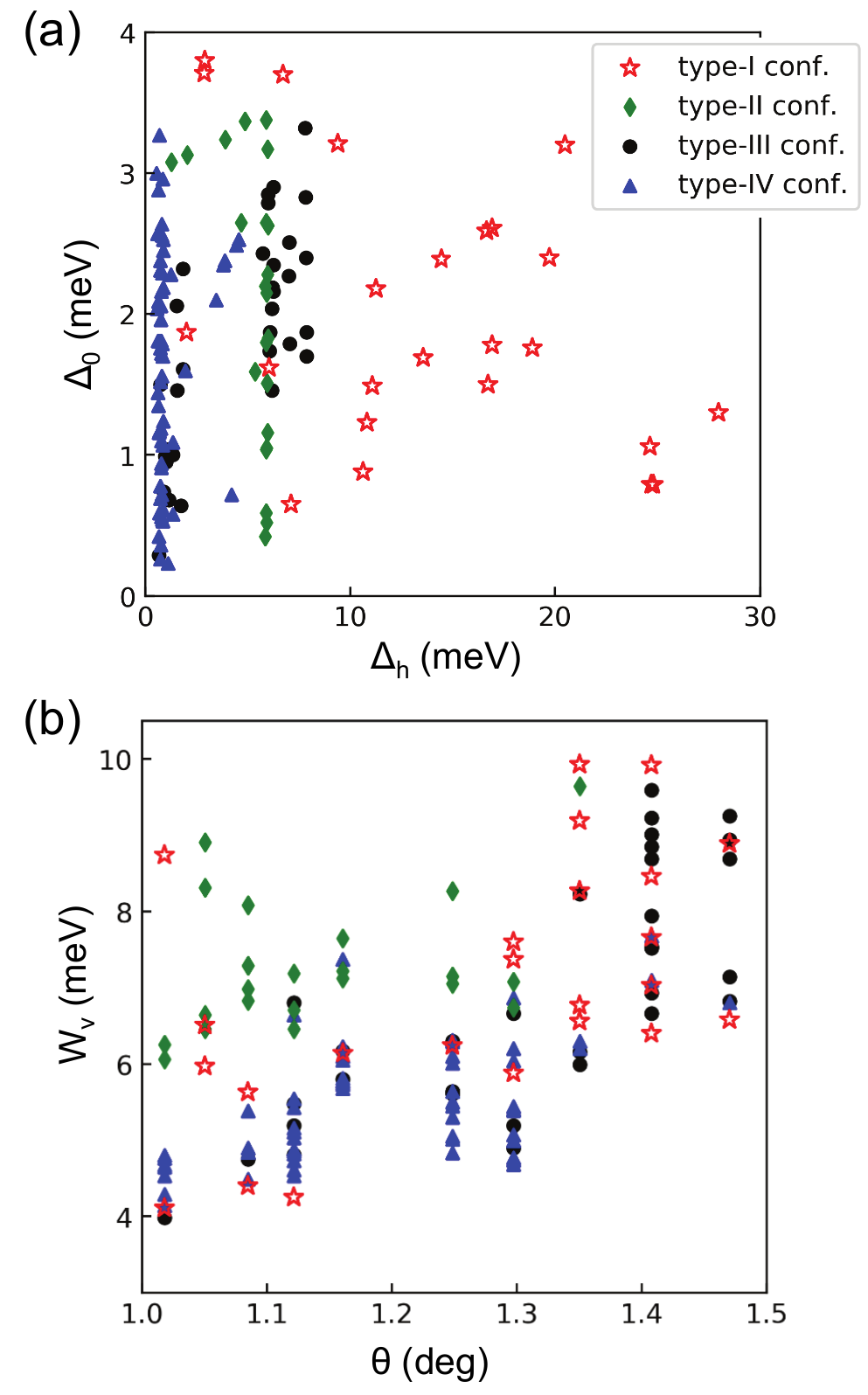}
\end{center}
\caption{(Color online)
The properties of isolated flat valence bands in configurations with different stacking types.
(a) $\Delta_0$ and $\Delta_h$ for all configurations with isolated flat valence bands.
Each configuration of TMLG is characterized by the stacking and $\theta$.
(b) Width ($W_v$) of the isolated valence band versus $\theta$ of these configurations.
\label{fig2}}
\end{figure}

\section{Structural configurations of TMLG}

\begin{figure*}[t]
\begin{center}
\includegraphics[width=1.9\columnwidth]{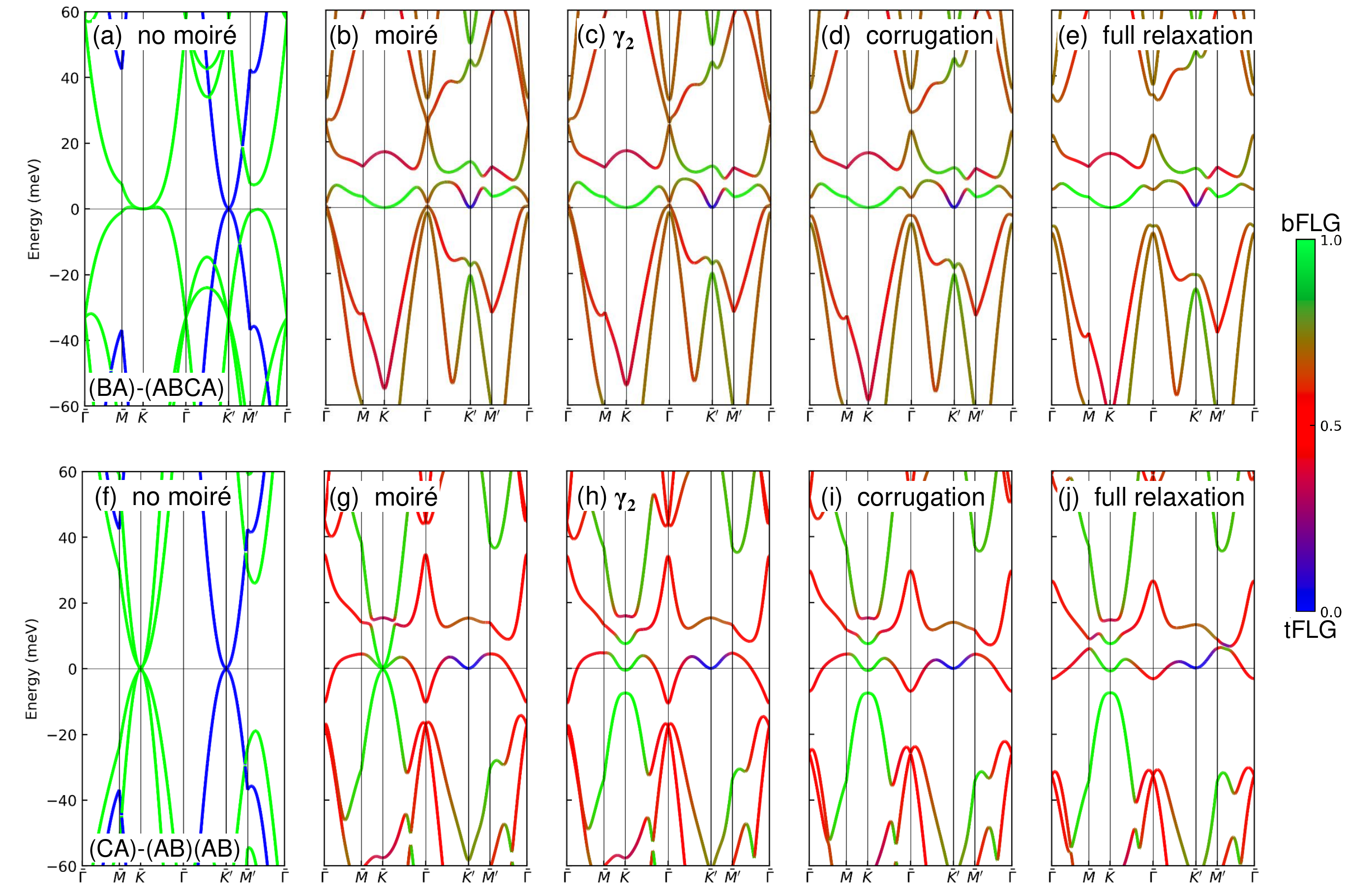}
\end{center}
\caption{(Color online) The band structures and state composition
for TMLG with the (BA)-(ABCA) stacking (a-e) and the (CA)-(AB)(AB) stacking (f-j) at $\theta = 1.35^\circ$.
Starting from the rigid superlattice without moir\'{e} coupling between the FLGs and without the $\gamma_2$ hopping in
the FLGs (a, e), the moir\'{e} coupling (b, g), the $\gamma_2$ hopping terms (c, h), the corrugation effect (d, i), and the in-plane relaxation effect (e, j)
are included in the Hamiltonian successively.
The band energies are directly calculated by diagonalizing the Hamiltonian without energy shifting,
so the Fermi levels do not necessarily lie at the zero energy here.
\label{fig3}}
\end{figure*}

\begin{figure}[t]
\begin{center}
\includegraphics[width=1.0\columnwidth]{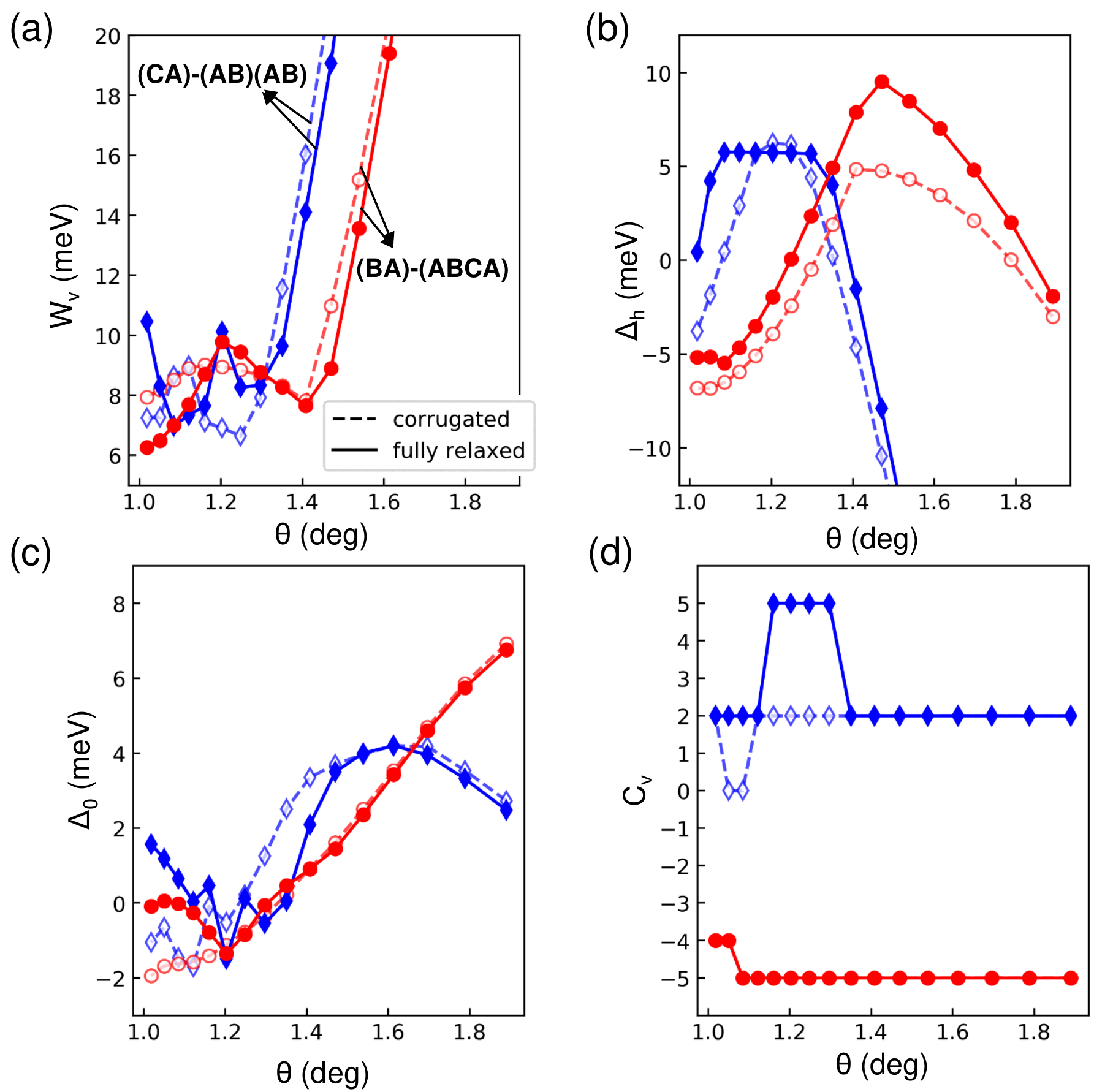}
\end{center}
\caption{(Color online)
Variations of $W_v$ (a), $\Delta_h$ (b), $\Delta_0$ (c), and the Chern number ($C_v$) of the highest valence band (d) with $\theta$
for the (BA)-(ABCA) and (CA)-(AB)(AB) stackings considering full relaxation (solid symbols) or only corrugation (open symbols).
\label{fig4}}
\end{figure}

\begin{figure}[b]
\begin{center}
\includegraphics[width=1.0\columnwidth]{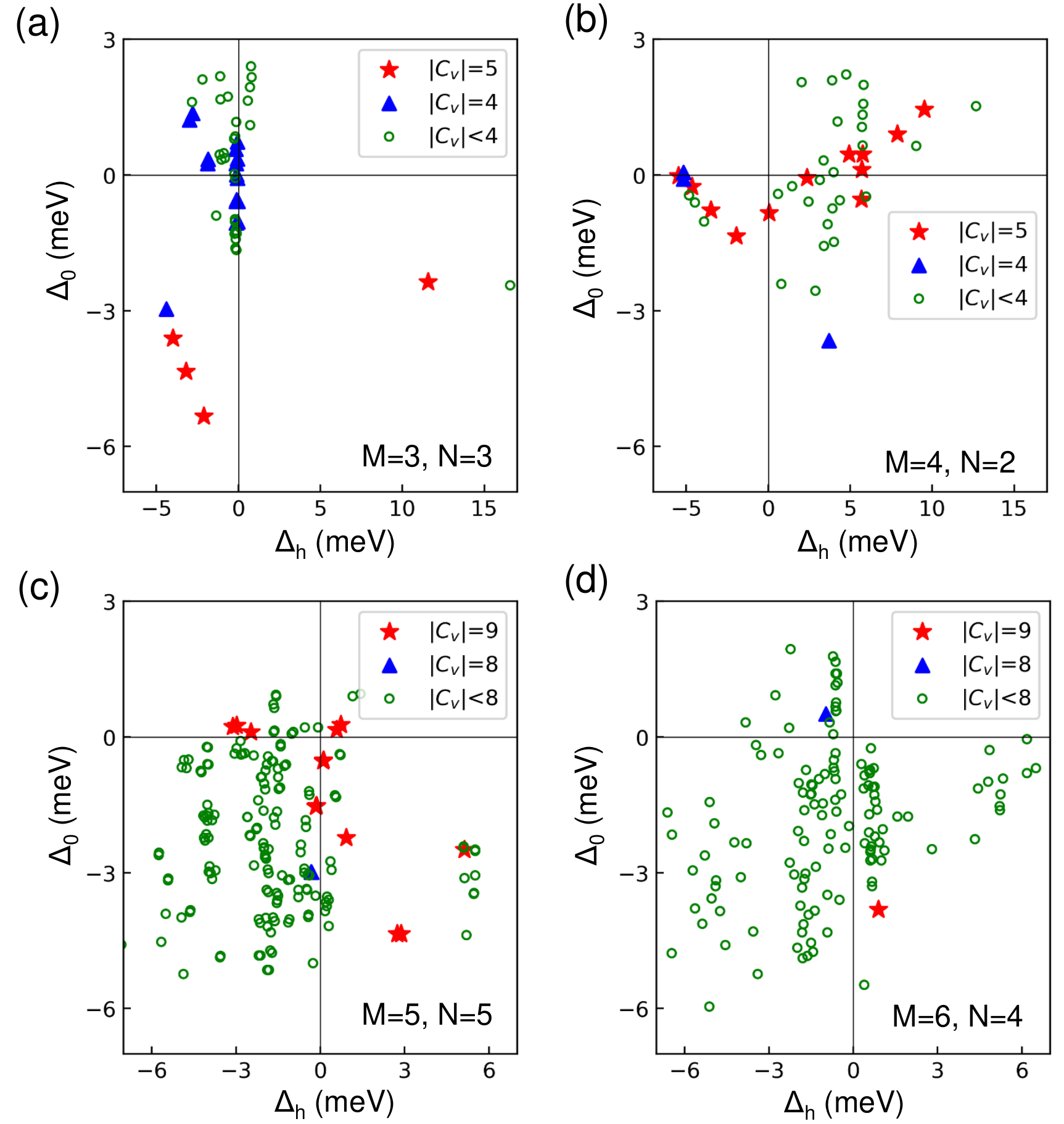}
\end{center}
\caption{(Color online)
The $|C_v|$,  $\Delta_0$ and $\Delta_h$ of configurations with separable flat valence bands at given $M$ and $N$.
$M+N=6$ in (a-b) and $M+N=10$ in (c-d).
\label{fig5}}
\end{figure}

We study TMLG with the top FLG (tFLG) rotated
by $\theta$ counterclockwise and the bottom FLG (bFLG) fixed, as seen in Fig. 1(a).
The layer numbers of bFLG and tFLG are denoted by $M$ and $N$, respectively.
We consider the strictly periodic moir\'{e} superlattices of TMLG with $M + N$ up to 10 and $\theta$ from 1.890$^\circ$ to 1.018$^\circ$.
Starting from the twist interface, the layers in bFLG and tFLG are indexed by $n = 1 \sim M$ and
$\tilde{n} = \tilde{1} \sim \tilde{N}$, respectively.
Besides the tunable $\theta$,
there exist $2^{M+N-3}$ inequivalent stacking orders for TMLG with $M+N$ layers,
which are represented by the stackings of tFLG and bFLG. For example,
a configuration composed of tFLG with the BA stacking and bFLG with the ABCA stacking is denoted
by BA-ABCA. The geometry of moir\'{e} superlattices in TMLG is detailed in the Supplemental Material
(SM).

Within a FLG, a pair of sites from two adjacent layers with one site directly above the other are referred to as dimer sites,
and there is a relatively strong interlayer hopping ($\gamma_1$) between them.
In a chirally stacked FLG, the sites in inner layers are all dimer sites, and the non-dimer sites which constitute the low-energy band
states are on the surface layers\cite{Trigonal2009Koshino,Band2010zhang}.
For a general stacking of FLG, the layers can be decomposed into a sequence of chiral subsets
\cite{Chiral-decomposition-2008,min2008electronic,zhang2020chiral,cao2020textit},
which are represented as chiral stacking orders in parentheses, such as (AB)(ABC). Such chiral decomposition
of tFLG and bFLG in TMLG determines some characters of the electronic structure.

In TMLG, the local stacking between layer $\tilde{1}$ of tFLG and layer $1$ of bFLG varies continuously across the moir\'{e} superlattice.
Then the optimal local spacing between these layers determined by the local stacking also varies with the in-plane position in the superlattice,
leading to the corrugation of the layers. More importantly, the spatially varying potential at the twist interface
can drive the in-plane structural relaxation. Full relaxation has been performed for each configuration of TMLG employing the continuum elastic theory,
as detailed in the SM.

\section{Electronic structure of TMLG}

\begin{table*}[t]
\caption{
The stacking orders that host isolated flat valence bands in a relatively large range of $\theta$:
the total layer numbers ($M+N$), the layer numbers of bFLG ($M$) and tFLG ($N$), the stacking type,
the twist-angle range from $\theta_1$ to $\theta_2$ with isolated flat valence bands, the values of
$W_v$, $\Delta_h$, and $\Delta_0$ in units of meV at $\theta_1$ and $\theta_2$, and the Chern numbers ($C_v$) of the flat valence bands
for twist angle from $\theta_1$ to $\theta_2$.
}
\begin{tabular*}{1.0\textwidth}{@{\extracolsep{\fill} }cccccccccccccccccc}
\hline\hline
$M+N$ & $M$ & $N$ & stacking & type & $\theta_{1}$ & $W_v^{(1)}$ & $\Delta_h^{(1)}$ & $\Delta_0^{(1)}$ &$\theta_{2}$ & $W_v^{(2)}$ & $\Delta_h^{(2)}$ & $\Delta_0^{(2)}$ & $C_v$\\
\hline
4 & 2 & 2 & (BA)-(AB) & I & 1.35$^\circ$ & 9.9 & 23.9 & 0.8 & 1.02$^\circ$ & 4.1 & 5.6 & 2.7 & -3,0,-2\\
5 & 3 & 2 & (BA)-(A)(BA) & IV & 1.30$^\circ$ & 5.4 & 0.8 & 1.3 & 1.02$^\circ$ & 4.7 & 0.7 & 2.0 & 0,-3,-2\\
6 & 3 & 3 & (AB)(A)-(A)(BA) & IV & 1.30$^\circ$ & 5.4 & 0.6 & 1.6 & 1.02$^\circ$ & 4.5 & 0.8 & 2.2 & 0,-3,-2\\
6 & 4 & 2 & (BA)-(AB)(AB) & II & 1.30$^\circ$ & 6.7 & 5.7 & 1.1 & 1.02$^\circ$ & 6.1 & 4.8 & 2.2 & 2,-1,0\\
6 & 4 & 2 & (CA)-(AB)(AB) & II & 1.35$^\circ$ & 9.6 & 4.0 & 0.1 & 1.05$^\circ$ & 8.3 & 4.2 & 1.2 & 2,5\\
6 & 4 & 2 & (BA)-(ABCA) & I & 1.47$^\circ$ & 8.9 & 9.5 & 1.4 & 1.35$^\circ$ & 8.3 & 4.9 & 0.5 & -5\\
7 & 5 & 2 & (BA)-(AB)(ABC) & III & 1.30$^\circ$ & 4.9 & 1.4 & 1.4 & 1.02$^\circ$ & 4.0 & 0.3 & 0.1 & -3,-6,-5\\
7 & 5 & 2 & (CA)-(AB)(ABC) & III & 1.25$^\circ$ & 6.3 & 1.4 & 0.4 & 1.08$^\circ$ & 4.8 & 0.5 & 0.5 & 0\\
8 & 4 & 4 & (BA)(BA)-(AB)(AB) & II & 1.30$^\circ$ & 7.1 & 5.7 & 0.5 & 1.02$^\circ$ & 6.2 & 4.9 & 2.2 & 4,1,2\\
8 & 5 & 3 & (AB)(A)-(AB)(ABC) & IV & 1.30$^\circ$ & 4.8 & 0.8 & 1.7 & 1.12$^\circ$ & 5.0 & 0.7 & 0.5 & -3,-6\\
8 & 6 & 2 & (BA)-(A)(BA)(BAC) & IV & 1.30$^\circ$ & 4.7 & 0.8 & 1.6 & 1.02$^\circ$ & 4.1 & 0.5 & 1.9 & 3,0,1\\
9 & 6 & 3 & (AB)(A)-(A)(BA)(BAC) & IV & 1.30$^\circ$ & 4.7 & 0.7 & 1.8 & 1.02$^\circ$ & 4.7 & 0.6 & 1.5 & 3,0,1\\
9 & 7 & 2 & (BA)-(A)(BA)(CABC) & IV & 1.25$^\circ$ & 6.1 & 1.1 & 0.7 & 1.02$^\circ$ & 4.8 & 0.6 & 0.1 & -4,-7,-6\\
10 & 5 & 5 & (CBA)(BA)-(AB)(ABC) & III & 1.30$^\circ$ & 5.2 & 1.4 & 0.9 & 1.12$^\circ$ & 5.5 & 0.6 & 0.2 & -6,-9\\
10 & 7 & 3 & (AB)(A)-(A)(BA)(CABC) & IV & 1.30$^\circ$ & 5.4 & 0.8 & 0.9 & 1.02$^\circ$ & 4.8 & 0.5 & 0.1 & -4,-7,-6\\
\hline\hline
\end{tabular*}
\end{table*}

\begin{figure*}[t]
\begin{center}
\includegraphics[width=2.0\columnwidth]{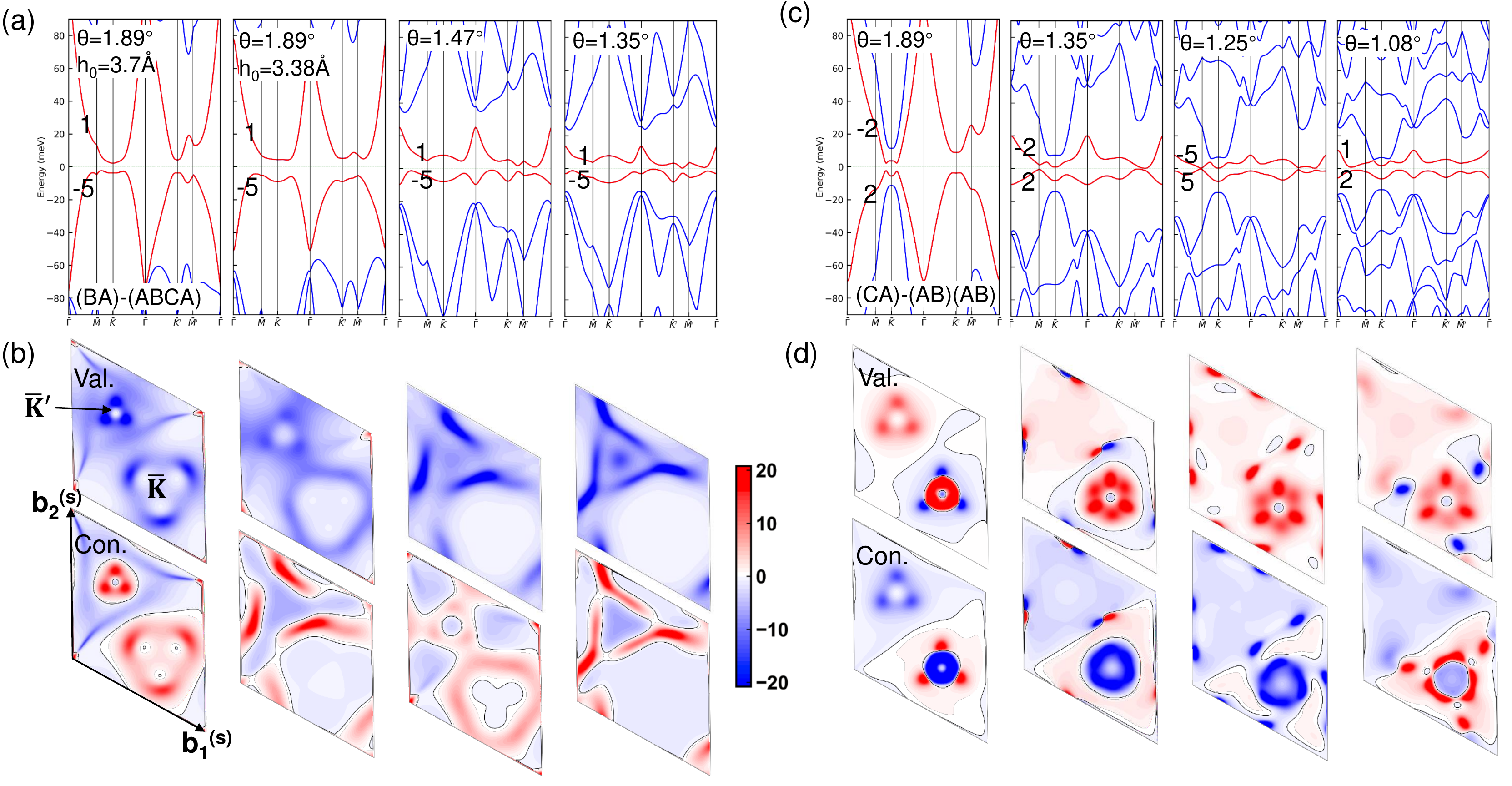}
\end{center}
\caption{(Color online) The band structures and $\Omega_z$ maps of the middle valence and conduction bands around $E_F$ for the (BA)-(ABCA) (a-b)
and the (BA)-(ABCA) (c-d) stackings with different $\theta$ and $h_0$.
In (a, c), the middle bands are displayed in red and their Chern numbers are labeled,
and the Fermi levels are represented by dashed lines.
In (b, d), the $\Omega_z$ maps are illustrated in the supercell BZ and the zero level of $\Omega_z$ is represented by gray
contour lines.
\label{fig6}}
\end{figure*}

\begin{figure}[t]
\begin{center}
\includegraphics[width=0.8\columnwidth]{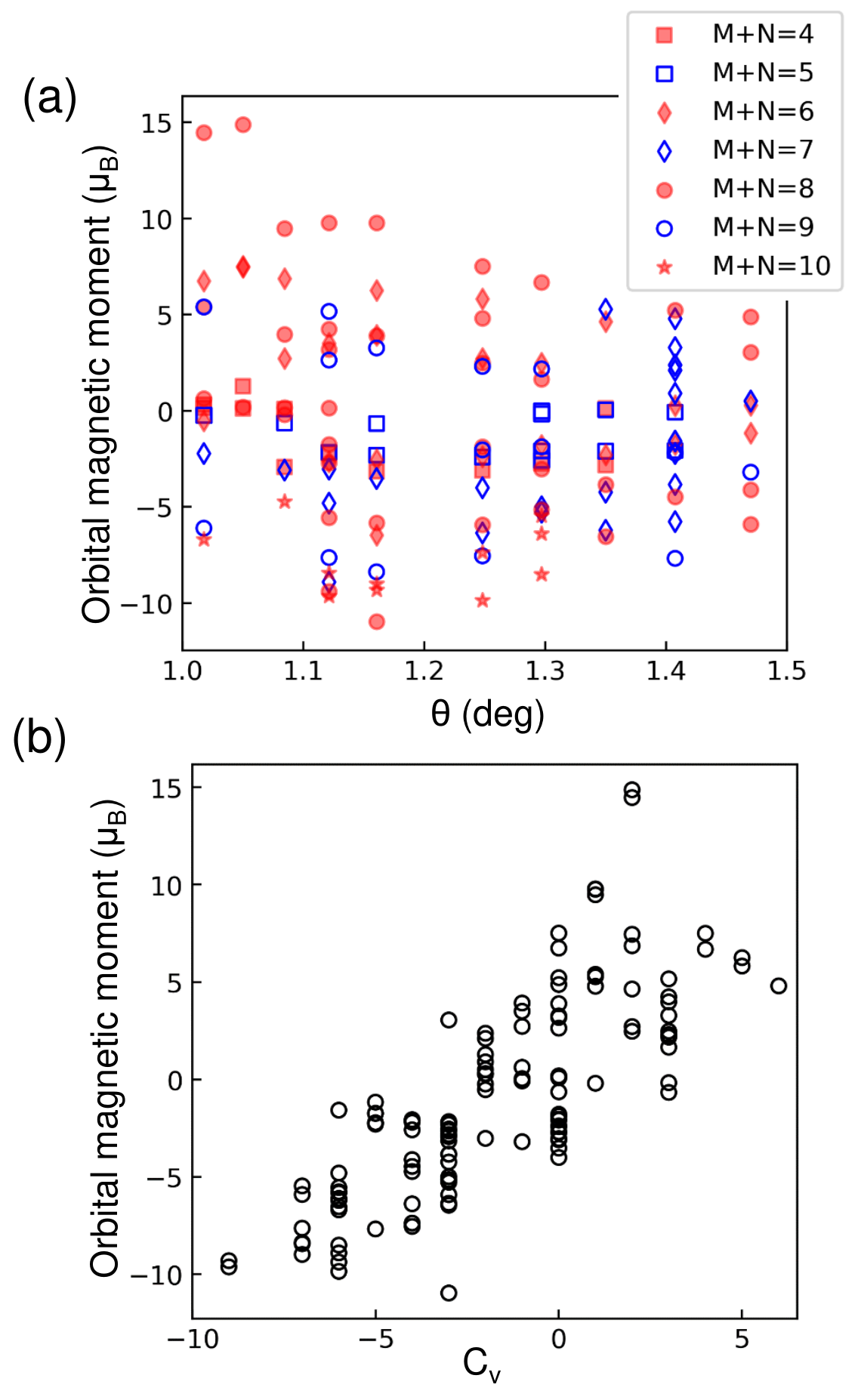}
\end{center}
\caption{(Color online)
(a) The orbital magnetic moments ($m$)
contributed by the isolated flat valence band for configurations with different layer numbers ($M+N$).
(b) $m$ versus $C_v$ for all configurations with isolated flat valence bands.
\label{fig7}}
\end{figure}

For each relaxed configuration of TMLG, we have built a tight-binding Hamiltonian ($\hat{H}$) taking into account the effects of
out-of-plane corrugation and in-plane structural deformation. The Hamiltonian
parameters and computational approaches of the electronic structure are given in the SM.
It is noted that the vertical next-nearest-layer hopping ($\gamma_2$) within the top and bottom FLGs is included in $\hat{H}$.
To diagonalize $\hat{H}$, the plane-wave-like basis functions are adopted and denoted by
$|n\alpha, \mb{k}\rangle$, where $n$ is the layer index, $\alpha = A, B$, and
$\mb{k}$ represents the momentum of the state in the reciprocal space of the pristine FLGs.

By calculating the two-dimensional ($2D$) energy bands in the entire Brillouin zone (BZ) for relaxed TMLG with different $\theta$ and stackings, we find that
completely isolated flat bands emerge around $E_F$ for various configurations of TMLG, even for those with 10 layers.
We consider flat bands with widths smaller than 10 meV.
For an isolated band, global gaps are opened above and below this band.
The bands around $E_F$ show rather strong electron-hole asymmetry and most isolated flat bands are on
the hole side.
The stacking orders that host isolated flat valence bands in a relatively large range of $\theta$ are listed in Table I
and those with flat valence bands that are only isolated at some $\theta$ can be seen in Table SI of the SM.
Figure 1(b-e) show the band structures of four typical configurations with different types of stacking orders and their DOS can be seen in Fig. S3.
Since only very few configurations have isolated flat conduction bands (see Table SII), we focus on isolated flat valence bands in the following.

Through examination of variations of the electronic properties with different configurations of TMLG,
we category the stacking orders into four types, i.e. cases with chirally stacked FLGs (type I), those with
Bernal-stacked FLGs that have even numbers of layers (type II), other cases without a single layer in the stacking decomposition of FLGs (type III)
and with a single layer in decompositions (type IV), as listed in Tables I and SI.
The isolated flat valence band is separated from other bands by global gaps at the charge neutrality point ($\Delta_0$)
and just below it ($\Delta_h$).
$\Delta_0$ is smaller than 4 meV for all cases, while $\Delta_h$
can have a rather large value, especially for type-I cases, as shown in Fig. 2(a).
Most type-IV configurations have extremely narrow $\Delta_h$ as some characters of the low-energy linear dispersions contributed by
the decomposed single layer in the pristine FLGs are maintained.
The $\Delta_h$ of most type-II cases are quite large, while those of many type-III cases are rather small.
The widths ($W_v$) of isolated valence bands begin to become smaller than 10 meV at $\theta \approx 1.5^\circ$,
as shown in Fig. 2(b).
For type-I cases, $W_v$ can be narrower than 5 meV only when $\theta$ is around or below $1.1^\circ$,
while the $W_v$ of type-III and type-IV cases can reach such narrow $W_v$ at about $1.3^\circ$.
The electronic behavior of different stacking types also depends on the thickness of TMLG.
Only when $N+M \leq 7$, isolated flat bands can emerge for type-I stacking, as seen in Table II.
For $N+M \geq 9$, only type-III and type-IV cases can have isolated flat bands.
No isolated flat bands exist for systems with $N = 1$.
In addition, similar to pristine FLGs\cite{Trigonal2009Koshino,Band2010zhang}, the 2D energy dispersions of the isolated flat bands exhibit trigonal
warping, as seen in Fig. S2.

To reveal the mechanism behind the emergence of isolated flat bands in TMLG, we have compared the band structures and state composition of
some configurations with different parts of the Hamiltonian, as shown in Fig. 3
for the type-I (BA)-(ABCA) stacking and the type-II (CA)-(AB)(AB) stacking  at $\theta = 1.35^\circ$.
Starting from the rigid superlattice without moir\'{e} coupling between FLGs and without the $\gamma_2$ hopping within
FLGs, the moir\'{e} coupling, the $\gamma_2$ hopping, the corrugation effect, and the in-plane relaxation effect are included in the Hamiltonian
successively.

Without moir\'{e} coupling, the low-energy bands of bFLG (tFLG) touch at the corner $\bar{K}$ ($\bar{K}'$) [see Fig. S1(b)] of the supercell BZ,
and the bands of bFLG cross with those of tFLG, as seen in Figs. 4(a), 4(f) and S4. The band crossings can be eliminated by the moir\'{e}
hopping between layers 1 and $\tilde{1}$ in a way similar to TBG, as these degenerate states at crossings have contributions from these interface layers.
In contrast to TBG always with touched valence and conduction bands at the BZ corners, state splitting occurs around $E_F$ upon
inclusion of the moir\'{e} coupling for TMLG.
In the pristine bFLG, the states composed of non-dimer sites have zero energy at $\bar{K}$.
These states are just the non-dimer basis functions with the momentum of $K_+$, which is at the corner of the bFLG BZ [see Fig. S1(a)].
When bFLG is coupled to tFLG, the energy of the non-dimer state ($|1B, K_+\rangle$) in the first layer of bFLG
rises above zero and the other non-dimer states still have zero energy.
We will show that
the positive energy of $|1B, K_+\rangle$ is due to the electron-hole asymmetry in low-energy bands of
tFLG.

For the chiral tFLG whose zero-energy states are at $\bar{K}'$, three degenerate conduction states
($|\psi_{c,n}\rangle$ with $n = 1-3$) and three degenerate valence states ($|\psi_{v,n}\rangle$) have low energies ($\varepsilon_{c, n}$ and
$\varepsilon_{v, n}$) at $\bar{K}$, as seen in Fig. S4.
By second-order perturbation approximation, the energy ($\varepsilon_{1B}$) of $|1B, K_+\rangle$ can be expressed as
\begin{equation}
\varepsilon_{1B} = \sum_{n=1}^{3} \frac{|\langle 1B, K_+ | \hat{H} |\psi_{c,n} \rangle|^2}{-\varepsilon_{c, n}} +
                    \sum_{n=1}^{3} \frac{|\langle 1B, K_+ | \hat{H} |\psi_{v,n} \rangle|^2}{-\varepsilon_{v, n}}
\end{equation}

Considering only the interlayer hopping $\gamma_1$ between dimer sites, the bands of tFLG are
electron-hole symmetric. Then all $|\langle 1B, K_+ | \hat{H} |\psi_{c,n} \rangle|$ and $|\langle 1B, K_+ | \hat{H} |\psi_{v,n} \rangle|$
have the same value. One of $|\psi_{c,n}\rangle$ is expanded in the basis functions with momentum at $K_+$ including $|\tilde{1}\alpha, K_+\rangle$ ($\alpha=A, B$)
in layer $\tilde{1}$ of tFLG.
For the rigid superlattice, the moir\'{e} coupling between layers $1$ and $\tilde{1}$
produces the same value ($u$) of the Hamiltonian elements between $|1B, K_+\rangle$ and $|\tilde{1}\alpha, K_+\rangle$.
For the BA stacked tFLG, $|\langle 1B, K_+ | \hat{H} |\psi_{c,n} \rangle|$ becomes $u/\sqrt{2}$ as layer $\tilde{1}$
contributes half of the squared norm of $|\psi_{c,n}\rangle$. For other tFLG, $|\langle 1B, K_+ | \hat{H} |\psi_{c,n} \rangle|$
decreases slowly with its number of layers.
With the electron-hole symmetry, $\varepsilon_{v, n} = -\varepsilon_{c, n}$, then $\varepsilon_{1B}$ is still zero.

The interlayer hopping $\gamma_4$ between a non-dimer site and a dimer site introduces electron-hole asymmetry in
the bands of FLGs. The supercell momentums of the basis functions expanding $|\psi_{v,n}\rangle$ and $|\psi_{c,n}\rangle$
all have the same length $k_\theta \approx 4\pi\theta/3a$.
With only $\gamma_1$ and $\gamma_4$ as well as the intralayer nearest-neighbor hopping (-$t_0$) for chiral tFLG with $N$ layers, $\varepsilon_{v, n}$ and $\varepsilon_{c, n}$
can be expressed analytically as\cite{Trigonal2009Koshino,Band2010zhang}
\begin{eqnarray}
\varepsilon_{v, n} &=& -(\sqrt{3}t_0 a k_\theta/2)^N/\gamma_1^{N-1} + 3 t_0 \gamma_4 a^2 k_\theta^2/(2 \gamma_1), \nonumber \\
\varepsilon_{c, n} &=& (\sqrt{3}t_0 a k_\theta/2)^N/\gamma_1^{N-1} + 3 t_0 \gamma_4 a^2 k_\theta^2/(2 \gamma_1),
\end{eqnarray}
which lead to $-\varepsilon_{v, n} < \varepsilon_{c, n}$.
Numerical calculations show that $\varepsilon_{v, n}$ is also closer to zero than $\varepsilon_{c, n}$
for non-chiral FLGs.
In addition, the norms of all the Hamiltonian elements in Eq. (1) are still
approximately equal when including $\gamma_4$. Therefore, $\varepsilon_{1B}$ has a positive value. For the BA stacked tFLG,
$\varepsilon_{1B}$ is given analytically by
\begin{equation}
\varepsilon_{1B}  = \frac{u^2}{\theta^2}\frac{9\gamma_1}{8\pi^2 t_0^2} (\frac{1}{1 - 2\gamma_4/t_0} - \frac{1}{1 + 2\gamma_4/t_0}).
\end{equation}
As $\gamma_4$ is much smaller than $t_0$, we have $\varepsilon_{1B} \approx 9\gamma_1 \gamma_4 u^2/(2\pi^2 t_0^3 \theta^2)$.

At $\bar{K}'$, similar analysis shows that
the positive energy of the non-dimer state $|\tilde{1}B, K_+'\rangle$ in layer $\tilde{1}$ of tFLG can also be attributed to the
electron-hole asymmetry in the bands of bFLG.

When only moir\'{e} coupling is included in the Hamiltonian of the (CA)-(AB)(AB) configuration with non-chiral bFLG,
three states contributed by the non-dimer sites in the bottom three layers of bFLG
still have zero energy at $\bar{K}'$, as seen in Fig. 3(g). The turning on of the $\gamma_2$ hopping between
the 2A and 4A sites can split these degenerate states, as shown in Fig. 3(h). Then only the
$|3B, K_+'\rangle$ state is located at almost zero energy.

For the rigid superlattice with all interlayer hopping, the flat bands around $E_F$ remain overlapped with other dispersive bands.
The corrugation effect reduces such band overlapping by making the flat valence band narrower, as shown in Figs. 3(d) and 3(i).
The in-plane relaxation has a more significant impact on the band dispersions. The flat valence band can become completely gapped from other bands
with full relaxation, while $\Delta_0$ may be decreased by the in-plane relaxation, especially for configurations with a non-chiral FLG, as shown in
Figs. 3(e) and 3(j).

To further demonstrate the importance of full relaxation for the emergence of isolated flat bands,
the variations of $W_v$, $\Delta_h$ and $\Delta_0$ with $\theta$ considering both out-of-plane and
in-plane relaxation are compared with those considering only corrugation in Fig. 4.
For (BA)-(ABCA) with only corrugation effect, $W_v$ is overestimated at $\theta$ around 1.1$^\circ$, $\Delta_h$ is underestimated at all $\theta$,
and $\Delta_0$ is also underestimated  at small $\theta$.
For (CA)-(AB)(AB), only with full relaxation can $\Delta_h$ and $\Delta_0$ become positive at small $\theta$.
The trends of these electronic properties with $\theta$ are also rather different for the two stackings. In particular,
the maximum value of $\Delta_h$ is located at about 1.5$^\circ$ for (BA)-(ABCA), while it is at about 1.2 $^\circ$ for (CA)-(AB)(AB).
In addition, the first local minimum of $W_v$ is also at a larger $\theta$ for (BA)-(ABCA).

We have focused on TMLG without potential differences between layers.
It is noted that equal potential differences ($\Delta$) between adjacent layers produced by applied out-of-plane electric field
may separate crossed bands locally for some configurations, while only a rather small $\Delta$ can enhance the gaps around
the isolated flat bands slightly and a larger $\Delta$ tends to close these global gaps, as shown for example in Fig. S5
for the (BA)-(ABCA) and (BA)-(ABCA) configurations.

\section{Valley Chern numbers and orbital magnetic moments}

At an odd filling of an isolated flat band,
spontaneous valley polarization may occur due to the electron-electron interaction\cite{Nearly-flat-2019}.
If such a valley polarized band hosts a non-zero Chern number, the TMLG can support QAHC.
The Chern numbers $C$ of flat bands in the $\xi = +$ valley have been obtained explicitly by
integral of the Berry curvature ($\Omega_z$) in the supercell BZ as detailed in the SM, and
the $C$ of a band in the $\xi = -$ valley is just
the opposite of that for $\xi = +$.
Chern numbers are also calculated for separable flat bands which are separated from nearby bands by
local gaps larger than 0.5 meV to characterize the dependence of $C$ on stacking orders of TMLG.

We also focus on the Chern numbers ($C_v$) of flat valence bands in the $\xi = +$ valley.
Among all configurations with $M+N$ layers, the largest $|C_v|$ is $M+N-1$.
The largest $|C_v|$ can occur in certain configurations with isolated flat bands, such as those shown in Figs. 1(b) and 1(d) and listed in Table SI.
We note that the isolated flat valence band in a configuration with 10 layers [see Fig. 1(d)] has $C_v = -9$,
which is the largest magnitude of $C_v$ for all considered TMLG with $N+M\leq10$.

Systematic calculations of all configurations with different stackings and $\theta$
show that most flat valence bands with high $C_v$ slightly overlap with
other bands and such band overlapping is related to the layers numbers ($M$ and $N$) of FLGs in TMLG, as illustrated in Fig. 5.
The appearance of highest $C_v$ generally becomes less likely with increasing N+M.
For $N+M=6$, more cases with $|C_v|=5$ appear for $M=4$ than for $M=3$ as type-II stacking is only possible for $M=4$.
In contrast, there are much fewer cases with the largest $|C_v|$ for $M=6$ than for $M=5$ when $M+N=10$, which is related to
the occurrence of type-III configurations only for $M=5$.
We notice that it is mainly the overlapping of the valence and conduction bands around EF, i.e. negative $\Delta_0$ that
keeps most flat valence bands from being isolated for large $N+M$.

By examining the evolution of $\Omega_z$ maps for the middle bands with the strength of the
moir\'{e} coupling, we find that the large $|C_v|$ originates from the splitting of degenerate states
by the weak moir\'{e} coupling or from the topological transitions induced by the strong
coupling.
With the decreasing $\theta$ the coupling between FLGs can be tuned from the weak to the strong regime.
At a fixed $\theta$, the coupling strength can be further reduced by rigidly increasing the average spacing ($h_0$) between FLGs.

Figure 6 displays the $\Omega_z$ maps of the valence and conduction bands around $E_F$ for the (BA)-(ABCA)
and (BA)-(ABCA) stacked configurations with different $\theta$ and $h_0$.
For (BA)-(ABCA) with $M=4$ and $N=2$,
$\Omega_z$ peaks appear around $\bar{K}$ and $\bar{K}'$ in the weak coupling regime.
The highest valence band states around $\bar{K}$ ($\bar{K}'$) are mainly contributed by the chiral bFLG (tFLG),
whose $\Omega_z$ is integrated to be about $-M\pi$ ($-N\pi$)\cite{Trigonal2009Koshino,Band2010zhang}.
Such integral values of $\Omega_z$ correspond to the Berry phases of the band states in a circle around a corner of the hexagonal BZ
for the pristine FLGs, and the sign of the Berry phases is fixed by the fact that the valence state at $\bar{K}$ ($\bar{K}'$)
is composed of non-dimer states in
the bottom layer of bFLG (the top layer of tFLG) as discussed above.
These Berry phases around $\bar{K}$ and $\bar{K}'$ contribute $-(M+N)/2 = -3$ to the total $C_v$ for (BA)-(ABCA), and
the Chern number ($C_c$) of the lowest conduction band gains a contribution of $3$.
Moreover, the degeneracy lifting at the band crossings between the middle bands and the other bands
gives rise to negative $\Omega_z$ peaks along three curves beginning from the $\bar{\Gamma}$ point for both the valence and conduction bands,
and both $C_v$ and $C_c$ gain a contribution of -2 from these $\Omega_z$ peaks.
Then $C_v$ and $C_c$ amount to -5 and 1, respectively.
Similar to the 2D band dispersions, the distributions of $\Omega_z$ around $\bar{K}$ and $\bar{K}'$ exhibit trigonal warping inherited from
the pristine chiral FLGs.
With the decreasing $\theta$ or $h_0$, the feathers of the $\Omega_z$ maps undergo great changes with the peaks merging, while
the local gaps between the middle bands remain opened and the Chern numbers are maintained in a large range of $\theta$, as seen in Fig. 4(d).
Such evolution of $\Omega_z$ maps suggests that exploration of the electronic structure of TMLG in the weak regime of the moire coupling is essential
to interpret the origin of the Chern numbers at small $\theta$.
For some configurations with non-chiral FLGs, the largest $|C_v|$ can also be already present at large $\theta$, such as
the (BA)-(ABCA)(CB) stacking with $C_v = -7$ [see Fig. S6(a)]
for $\theta$ from 1.89$^\circ$ to 1.47$^\circ$ at which the flat valence band becomes isolated.

In the weak coupling regime for (CA)-(AB)(AB), the $\Omega_z$ peaks around $\bar{K}'$ originating from tFLG contribute +1
to $C_v$, while there are both positive and negative peaks around $\bar{K}$ as bFLG is non-chiral, then
$C_v$ has a relatively small value of 2.
With decreasing of $\theta$, dipole-like pairs of positive and negative peaks form and band inversion between the middle bands occur at about 1.30$^\circ$,
leading to the largest $C_v$ of 5. The next topological phase transition occurs at about 1.16$^\circ$.
We note that full relaxation is required to drive the topological phase transition with $C_v$ rising to 5 in the strong coupling regime, as shown in Fig. 4(d).
The $C_v = -9$ for the (CBA)(BA)-(AB)(ABC) stacking shown in Fig. 1(e) also occurs at small $\theta$, and the variation of $\Omega_z$ maps with $\theta$
can be seen in Fig. S6(b).

The non-trivial topology in isolated flat bands suggests that spontaneous
orbital magnetic moments ($m$) may be observed at odd fillings of the flat bands\cite{Voltage-Controlled-2020,Ferromagnetism-in-2020}.
We evaluate the magnitude of $m$ contributed by the isolated flat valence band
by setting the chemical potential at the middle of the $\Delta_0$ gap.
The magnitude of $m$ can reach 10$\mu_B$ per supercell for $\theta$ around 1.1$^\circ$,
as shown in Fig. 7(a).
For configurations with positive $m$, its value is related to the layer numbers of TMLG with the
$N+M=8$ cases generally having the largest $m$ at a $\theta$.
$m$ is roughly proportional to $C_v$, while the spanning of its value is still rather large for a given $C_v$, especially
for small $|C_v|$.
We note that for certain cases with a small $|m|$ but a large $|C_v|$, the sign of the total orbital magnetic moment may be
inverted by tuning the chemical potential across the gap\cite{Voltage-Controlled-2020}.

\section{Summary and Conclusions}

For TMLG with $M+N$ layers, full relaxation has been performed for the $2^{M+N-3}$ inequivalent stacking orders at
varying $\theta$.
Isolated flat bands emerge in relaxed TMLG with $M+N$
up to 10
and with various stackings,
and most of them are on the hole side.
The stacking orders that host isolated flat bands can be categorized into four types based on the electronic behavior of flat bands and the stacking
decomposition.
For type-I configurations with both chiral FLGs,
the touched bands of FLGs around $E_F$
are split by the moir\'{e} coupling through the electron-hole asymmetry in low-energy bands of FLGs
while the hopping $\gamma_2$ between the next-nearest layers is further required for the state splitting in other types with non-chiral FLGs.
The corrugation effect reduces the band overlapping around $E_F$ and
the full structural relaxation leads to global gaps that completely isolate a flat band.
For TMLG with given $M$ and $N$, the highest Chern number $|C|$ of the separable flat bands
reaches $M+N-1$ and can be hosted by certain
isolated bands.
The $|C|=9$ occurs in the isolated flat valence band of several configurations with 10 layers.
Such high $|C|$ originates from the lifting of the band-state degeneracy in the weak regime of
moir\'{e} coupling or from the topological phase transitions induced by the strong
moir\'{e} coupling.
Moreover, large orbital magnetic moments $m$ arise in isolated flat bands with high $|C|$ and
depend on the structural configurations of TMLG.

\label{Acknowledgments}
\begin{acknowledgments}
We gratefully acknowledge valuable discussions with D. Tom\'anek,
H. Xiong, and S. Yin.
This research was supported by
the National Natural Science Foundation of China (Grants No. 11974312 and No. 11774195).
The calculations were performed on TianHe-1(A) at National Supercomputer Center in Tianjin.
\end{acknowledgments}


%

\end{document}